\pdfoutput=1
\documentclass{my-paper}

\title{The Bisq Decentralised Exchange: On the Privacy Cost of
  Participation\thanks{This is a preprint of three published
articles~\citep{hickey-harrigan-20,hickey-harrigan-21,hickey-harrigan-21-1}.}}

\author{Liam~Hickey\thanks{\email{liamhickeyire@gmail.com},
\orcid{0000-0002-4102-424X}}}

\author{Martin~Harrigan\thanks{\email{martin.harrigan@setu.ie},
\orcid{0000-0002-6069-7001}}}

\affil{\gls*{setu}, \acrlong*{roi}}

\date{}

\begin{document}

\maketitle

\begin{abstract}
  The Bisq Trade Protocol and the Bisq \gls*{dao} are core components of
Bisq, a decentralised cryptocurrency exchange.  The Bisq Trade
Protocol systematises the peer-to-peer trading of Bitcoin for other
currencies and the Bisq \gls*{dao} decentralises the governance and
finance functions of the entire exchange.  However, by following the
Bisq Trade Protocol and interacting with the Bisq \gls*{dao},
participants necessarily publish data to the Bitcoin blockchain and
broadcast additional data to the Bisq peer-to-peer network.  We
examine the privacy cost to participants in sharing this data.
Specifically, we use novel address clustering heuristics to construct
the one-to-many mappings from participants to addresses on the Bitcoin
blockchain and augment the address clusters with data stored within
the Bisq peer-to-peer network.  We describe address clustering
heuristics for both the Bisq Trade Protocol and the Bisq \gls*{dao}\@.
We show that the heuristics aggregate activity performed by each
participant: trading, voting, transfers, and so on.  We identify instances
where participants are operating under multiple aliases, some of which
are real-world names.  We identify the dominant transactors and their
role in a two-sided market.  We conclude with suggestions to better
protect the privacy of participants in the future.

\end{abstract}

\section{Introduction}\label{sec:introduction}

Bitcoin and its altcoin brethren, with the notable exception of
\textquote{privacy coins}, seek decentralisation first and privacy
second~\citep{harvey-branco-illodo-19}.  The pairing of
blockchain analysis service providers with regulated cryptocurrency
exchanges has exploited this.  The former perform blockchain-wide
analyses for high coverage but low individual identification.  The
latter enforce identity checkpoints for high individual identification
but low coverage.  Their pairing, combining aggregation with
identification, is an example of a well-known
privacy risk~\citep{solove-06}.

Bisq is a decentralised cryptocurrency exchange that does not enforce
identity checkpoints but relies on the Bitcoin blockchain and its own
peer-to-peer network to operate; thereby falling under the purview of
blockchain analysis service providers.  In this article we analyse the
Bisq Trade Protocol, the component of Bisq responsible for
systematising the peer-to-peer trading of Bitcoin for other
currencies, and the Bisq \gls*{dao}, the component of Bisq responsible
for decentralising the governance and finance functions of the entire
exchange, from a privacy perspective.  We show that there is a
significant privacy cost to participating in Bisq trades and the Bisq
\gls*{dao}\@.

Specifically, our analysis applies \emph{address clustering} with
Bisq-specific heuristics.  Address clustering is a cornerstone of
blockchain analysis.  It employs heuristics to partition the set of
addresses observed on a blockchain into \emph{address clusters} that
are likely controlled by the same participant.  When combined with
\emph{address tagging}, or associating real-world identities with
addresses, and graph analysis, it is an effective means of analysing
blockchain activity at both the micro- and macro-levels, see, for example,
\citep{reid-harrigan-13,meiklejohn-et-al-16,huang-et-al-18}.  The Bisq
Trade Protocol creates Bitcoin transactions with a distinctive
structure and is subject to this form of analysis.  Additionally, the
Bisq \gls*{dao} relies on a coloured-coin issued on the Bitcoin
blockchain known as the \unit{\bsq} token and is also subject to this
form of analysis.  We use the structure of the Bisq Trade Protocol
transactions and the structure of the \unit{\bsq} token transactions
to create Bisq-specific address clustering heuristics.

At the time of our analysis on 10th February 2021, or Bitcoin block
height \num{670026}, traders had completed approximately \num{90000}
trades using Bisq and the market capitalisation of the \unit{\bsq}
token was approximately \dUSD[prefix=million~]{13}.  Both numbers are
small when compared with the equivalent numbers for centralised
exchanges such as Coinbase and decentralised exchanges that deal only
with digital tokens such as Uniswap (see \cref{sec:related-work}).
However, Bisq is noteworthy and deserving of attention from a privacy
perspective for two reasons.  First, Bisq provides an on-ramp to the
world of cryptocurrencies without enforcing any identity checkpoints.
This could attract the scrutiny of regulators and blockchain analysis
service providers.  It is currently a blind spot in their coverage.
Indeed, recent guidance from the Financial Action Task Force (FATF)
regarding Virtual Asset Service Providers (VASPs) states that
\textquote[\citep{fatf-21}]{the decentralisation of any individual
  element of operations does not eliminate VASP coverage if the
elements of any part of the VASP definition remain in place.} If the
FATF considers Bisq to be a VASP, then the obligation to enforce
identity checkpoints may fall upon those who operate Bisq, even if the
operation is decentralised.  The ability to identify the contributors
will be central.  Second, Bisq is novel in its construction amongst
exchanges, in that it is decentralised \emph{and} it supports both
cryptocurrencies and fiat currencies.  It enables traders to exchange,
say, euros in a bank account for bitcoins, without having to trust a
counterparty or facilitator.  Although Bisq is the first of its kind,
it has spawned forks and competitors.  For example, \citet{haveno-xx}
is a fork of Bisq based on Monero instead of Bitcoin that cites our
earlier work~\citep{hickey-harrigan-20} as a motivating
factor.\footnote{\url{https://github.com/haveno-dex/haveno/wiki/FAQ}}
An analysis of privacy within Bisq can inform both blockchain analysts
and those seeking to hinder blockchain analysis.

This article is organised as follows.  In \cref{sec:related-work} we
review related work, including address clustering, decentralised
exchanges and decentralised governance.  In \cref{sec:bisq} we
introduce Bisq, the Bisq Trade Protocol, the Bisq \gls*{dao} and our
Bisq-specific address clustering heuristics.  We detail our analysis
and results in \cref{sec:analysis} and countermeasures in
\cref{sec:countermeasures}.  We conclude in
\cref{sec:conclusions}.

\section{Related Work}\label{sec:related-work}

We categorise related work into five areas: address clustering, token
analysis, transaction analysis, decentralised exchanges and
decentralised governance.

Address clustering is a fundamental building block upon which many
high-level blockchain analyses can be performed, see, for example,
\citep{reid-harrigan-13,meiklejohn-et-al-16,lischke-fabian-16,ermilov-et-al-17,filtz-et-al-17,di-francesco-maesa-et-al-18,huang-et-al-18,jourdan-et-al-18}.
Experimental analysis has shown that a single heuristic for address
clustering, the \emph{multi-input heuristic}, can identify more than
\qty{69}{\percent} of the addresses in the wallets stored by
lightweight clients~\citep{nick-15}.  This heuristic assumes that the
addresses referenced in transaction outputs spent in a single
multi-input transaction are controlled by the same
entity~\citep{nakamoto-08}.  Although vulnerable to techniques such as
CoinJoin~\citep{maxwell-13} and its kin, it is a useful heuristic in
practice~\citep{harrigan-fretter-16}.  Recently, specialised
approaches for sharing address tags~\citep{boshmaf-et-al-19},
standardising the scalable extraction and examination of blockchain
data~\citep{galici-et-al-20}, crowd-sourcing the classification of
transactions~\citep{weber-et-al-19}, parsing and representing the flow
of bitcoins between addresses~\citep{pinna-et-al-18}, and developing
address clustering heuristics for the Ethereum
blockchain~\citep{victor-20} have extended this line of research.

We use address clustering to track trade amounts, security deposits,
and the \unit{\bsq} token, a coloured-coin~\citep{rosenfeld-12} issued
on the Bitcoin blockchain by the Bisq project.  Tokens are a form of
\textquote{digital voucher} that provide access to a service or asset
while providing revenue or funding to token-based business
models~\citep{tasca-19}.  \citet{voshmgir-20} posits the token as the
primary building block of Web3 applications.  She surveys a variety of
token economies including those based on stable tokens, privacy
tokens, trading tokens, lending tokens, asset tokens, social tokens,
attention tokens and token curated registries.  She discusses the
technical, legal, economic and ethical aspects of token engineering.
There exist several network analyses of ERC-20 tokens on the Ethereum
blockchain that quantify their age, economic value and activity
volume~\citep{somin-et-al-18,victor-luders-19}.

Additionally, specialised heuristics have proved successful in tracing
transactions in \textquote{privacy coin} blockchains.  For example,
heuristics have been used to link public addresses on either side of
Zcash shielded transactions~\citep{quesnelle-17} and to identify the
true transaction inputs in Monero Ring Confidential
Transactions~\citep{moser-et-al-18}.

Decentralised exchanges enable traders to exchange cryptocurrencies
or fiat currencies without having to trust a centralised entity to
act as an intermediary for the exchange or as a custodian for the
currencies.  However, decentralised exchanges vary widely in terms of
technology, trustlessness and security~\citep{lin-19}.  Bisq is an
example of a decentralised exchange.  It goes to great lengths to
decentralise all aspects of its operation.  The terms
\emph{decentralised autonomous corporation} (DAC) and
\gls*{dao}~\citep{larimer-13,buterin-13,larimer-13-1} have a tainted
past due to the infamous failure of \textquote{The DAO} on the
Ethereum blockchain~\citep{dupont-17}.  However, progress continues
unabated~\citep{el-faqir-et-al-20}.  Decentralised exchanges are a
focus of DeFi (\emph{Decentralised Finance}).  DeFi includes several
projects that extend the decentralised nature of cryptocurrencies to
other areas of modern finance.  They are generally non-custodial,
permissionless, openly auditable and
composable~\citep{werner-et-al-21}.  DeFi projects typically take the
form of DApps (\emph{Decentralised Applications}), that operate using
smart contracts.  There are several decentralised exchange DApps, such
as Uniswap\footnote{\url{https://uniswap.org}},
SushiSwap\footnote{\url{https://www.sushi.com}},
0x\footnote{\url{https://0x.org}}, and many
others.\footnote{\url{https://distribuyed.github.io/index}} These
decentralised exchange DApps enable the exchange of ERC-20 tokens
using methods such as community powered liquidity pools or order book
based protocols.

Fully decentralised systems require decentralised governance.
\citet{reijers-et-al-18} distinguish between \emph{on-chain} and
\emph{off-chain} governance.  On-chain governance is a form of
decentralised governance that defers control to an underlying
blockchain.  Off-chain governance refers to all other rules and
decision-making processes that might affect the system.  The rules can
originate within the community that builds and maintains the system or
they can be imposed by external sources such as financial regulators.
There are tensions between both forms of
governance~\citep{bohme-et-al-15,lustig-nardi-15}.  Decentralised
exchanges use decentralised governance, in particular on-chain
governance, in a bid to circumvent rules imposed by external sources.
For example, they do not impose identity checkpoints or comply with
embargoes or watchlists.  It is unclear if fully decentralised
exchanges fall into the same regulatory perimeter as centralised ones,
and, if they do, how regulations are to be
enforced~\citep{de-filippi-14,aiyar-18}.

We use common terminology from graph theory throughout the article;
see \citet{diestel-17} for definitions.

\section{Bisq and Address Clustering}\label{sec:bisq}

The following is a description of Bisq, the Bisq Trade Protocol and
the Bisq \gls*{dao}; see \citet{bisq-xx,beams-karrer-17} for a more
thorough treatment.  We are particularly interested in the
transactions that are published by Bisq to the Bitcoin blockchain.  We
omit details regarding peer-to-peer messaging, the data stored locally
by Bisq nodes and the Bisq developer ecosystem.

Bisq, formerly known as Bitsquare, is a decentralised exchange that
enables traders to exchange bitcoins for altcoins (such as Ethereum,
Litecoin and Monero) and bitcoins for fiat currencies (such as USD,
EUR and GBP).  One side of each Bisq trade must involve bitcoins.
Bisq nodes connect to a peer-to-peer network over Tor to create an
order book, coordinate trades and resolve disputes.  Trades require
security deposits that are held using Bitcoin multi-signature
transactions.

A central tenet of Bisq, like Bitcoin, is the importance of
decentralisation.  Bisq has two goals: the first is to enable traders
to trade bitcoins for altcoins and fiat currency; the second is to
free traders from having to defer control to any central entity.  In
isolation, the first goal is easy to achieve.  Centralised exchanges
are a good example.  They are efficient and efficacious.  The caveat
is that centralised exchanges are subject to failures and regulatory
requirements.  Bisq achieves both goals but with some added costs.
For example, the user experience is arguably more complex than with a
centralised exchange and the system publishes transactions to the
Bitcoin blockchain incurring transaction fees and a privacy cost.

In the following we consider the privacy cost incurred by
participants.  Bisq is a double-edged sword for financial privacy.  It
allows anyone to trade with anyone else in the world without having to
divulge identifying information to a central entity.  However, it
requires participants to broadcast transactions on the Bitcoin
blockchain.  The contents of the transactions and their relationship
with other transactions can unintentionally disclose information about
the transactors to interested third parties.

\subsection{The Bisq Trade Protocol}\label{sec:bisq-trade-protocol}

\begin{figure}
  \myincludegraphics[width=0.9\linewidth]{tikz/blockchains/bisq-trade-protocol}
  \caption{A successful Bisq trade publishes four transactions to the
    Bitcoin blockchain.  The maker fee and taker fee transactions
    create transaction outputs that are redeemed by the deposit
    transaction.  The deposit transaction creates a multi-signature
  transaction output that is redeemed by the payout transaction.}
  \label{fig:bisq-trade-protocol}
\end{figure}

Bisq enables the non-custodial peer-to-peer exchange of bitcoins for
other currencies.  The Bisq Trade Protocol describes the sequence of
steps performed by traders when engaging in a trade.  The Bisq
software implements and executes the protocol on their behalf.  For
every successful trade, the Bisq software publishes four transactions
to the Bitcoin blockchain: a \emph{maker fee} transaction, a
\emph{taker fee} transaction, a \emph{deposit} transaction, and a
\emph{payout} transaction (see \cref{fig:bisq-trade-protocol}).  In
the maker fee transaction, the maker, or the trader adding liquidity
to the order book, pays a trade fee, a mining fee, a security deposit
and, if they are selling bitcoins, the trade amount.  In the taker fee
transaction, the taker, or the trader removing liquidity from the
order book, also pays a trade fee, a mining fee, a security deposit
and, if they are selling bitcoins, the trade amount.  Both traders
interact to create the deposit transaction: it combines the mining
fees, the security deposits and the trade amount from the maker fee
and taker fee transactions and locks the bitcoins in a multi-signature
transaction output.  The trader buying the bitcoins, or the buyer,
makes a payment to the seller using an altcoin or fiat currency.  The
traders interact to create the payout transaction: it refunds the
security deposits and sends the trade amount to the buyer.  There are
mechanisms in place to handle disputes, including mediation,
arbitration and time-locked transactions.

We can identify the four transactions in the Bitcoin blockchain for
successful Bisq trades.  The maker fee, taker fee, deposit and payout
transactions have a distinctive structure: every deposit transaction
has two transaction outputs, a transaction output whose redeem script
is a \num{2}-of-\num{2} or \num{2}-of-\num{3} multi-signature script,
and a transaction output with an \texttt{OP\_RETURN} containing the
hash of the Bisq trade contract.  The hash cannot be used to identify
Bisq trades.  However, the trade fee transaction outputs in the maker
fee and taker fee transactions are either sent directly to known Bisq
addresses that collect trade fees, or are coloured using the
\unit{\bsq} token (see \cref{sec:bsq} below).  We used the structure
of deposit transactions to identify \num{87326} Bisq
trades\footnote{\url{https://github.com/Liam-Hickey-Ire/BisqTradeProtocolAnalysisSource/blob/master/Data/670026/our-deposit-tx-hashes.csv}:
  Each line contains the hash of a deposit transaction that we
identified.} as of Bitcoin block height \num{670026}.

Bisq publishes and locally stores statistics relating to trades, and,
until recently, this included the complete list of deposit transaction
hashes.\footnote{As of Bisq~1.4.0, deposit transaction hashes are no
longer published and stored as part of the trade statistics.} We
cross-checked our list of deposit transaction hashes with the
\num{68517} trades stored by
Bisq\footnote{\url{https://github.com/Liam-Hickey-Ire/BisqTradeProtocolAnalysisSource/blob/master/Data/670026/bisq-deposit-tx-hashes.csv}:
  Each line contains the hash of a deposit transaction that Bisq
identified as part of the trade statistics.}.

\begin{table}
  \centering
  \caption{We can compare the deposit transactions identified by Bisq
    with the deposit transactions identified by us.  The two methods
  produce similar results.}\label{tab:deposit-txes}
  \begin{tabular}{llcc}
    \toprule
    \multicolumn{2}{c}{} & \multicolumn{2}{c}{Identified By Bisq in
    Trade Statistics}\\
    \cmidrule(lr){3-4}
    & & Deposit Tx & Not a Deposit Tx \\
    \midrule
    \multirow{2}{*}{Identified By Us} & Deposit Tx & \num{68048} & \num{541} \\
    & Not a Deposit Tx & \num{18} & \num{>300}~million \\
    \bottomrule
  \end{tabular}
\end{table}

We compared the two methods for the time-period between the first and
last deposit transactions that were identified and stored by Bisq as
part of its trade statistics.  \cref{tab:deposit-txes} shows that both
methods produced similar results.  They agreed on \num{68048} deposit
transactions; we identified \num{541} deposit transactions that were
not identified by Bisq; and Bisq identified \num{18} deposit
transactions that were not identified by us.  The differences were
small and, on inspection, were due in part to inaccuracies in the data
stored by Bisq.  This is an important finding since it shows that we
can continue to identify deposit transactions, and Bisq trades, even
though Bisq has stopped publishing and storing the deposit transaction
hashes as part of the trade statistics.

This leads to our Bisq Trade Protocol address clustering heuristic:
for each maker fee, taker fee, deposit and payout transaction
corresponding to a successful Bisq trade, the addresses referenced by
all of the transaction inputs of the fee transaction that does not
contain the trade amount, that is, the fee transaction created by the
trader buying bitcoins, and the address referenced by the transaction
output of the payout transaction containing the trade amount, belong
to the same trader.  The same applies in reverse to the trader selling
bitcoins.  Additionally, the addresses referenced by all of the
transaction inputs of the maker fee transaction and the address
referenced by the transaction output of the payout transaction
containing the maker's security deposit, belong to the maker, and
similarly for the taker fee transaction and the taker's security
deposit.  In other words, we can follow the trade amount and security
deposits through the four transactions.  The trade amount is exchanged
between the traders whereas the security deposits are returned to the
traders.  This allows the addresses referenced by the transaction
inputs of the maker fee and taker fee transactions to be clustered
with corresponding addresses referenced by the transaction outputs of
the payout transaction.  We can also add the addresses referenced by
the second and third output of the maker and taker fee transactions to
these clusters, the second output being an address used to construct
the deposit transaction controlled by the transaction's creator, and
the third output being a change output.  We exclude the first output
of the maker and taker fee transaction as this output contains the
Bisq trade fee itself.  \cref{app:trade-heuristic} provides
pseudo-code for the heuristic and we have released a C\#/.NET
implementation.\footnote{\url{https://github.com/Liam-Hickey-Ire/BisqTradeProtocolAnalysisSource}}

This heuristic shows that participating in a Bisq trade does not
obfuscate the flow of bitcoins between a seller and a buyer in any
way, and identifies the activity as a Bisq trade.  Of course, this
heuristic can be combined with the heuristics for clustering addresses
on the Bitcoin blockchain referenced in \cref{sec:related-work}.  We
will not expand upon this analysis in the remainder of this article
since its impact on privacy is evident.  Instead, we turn our
attention to the Bisq \gls*{dao} and a novel address clustering
heuristic based on its operation.

\subsection{The Bisq \gls*{dao}}\label{sec:bisq-dao}

There are two types of participant in the Bisq ecosystem: those who
use Bisq solely as a decentralised trading platform and those who take
part in the development, operation and governance of Bisq.  The Bisq
\gls*{dao} is the vehicle through which the latter group manages the
governance and finance functions of Bisq in a decentralised
fashion~\citep{beams-karrer-17}.  Participants in the Bisq \gls*{dao}
can make and vote upon proposals relating to Bisq using a stake-based
voting system.  They propose and vote on measures such as rewarding
contributors, approving support agents and modifying trade fees.
Voting occurs in approximately monthly cycles known as \gls*{dao}
cycles.  The \gls*{dao} cycle times are determined by block heights on
the Bitcoin blockchain.  As of Bitcoin block height \num{670026},
there have been twenty-one \gls*{dao} cycles.  Each cycle includes a
proposal phase, a blind vote phase, a vote reveal phase, and a vote
result phase.  The former group may also participate in the Bisq
\gls*{dao} to a lesser extent by acquiring and burning \unit{\bsq}
tokens in lieu of paying trading fees denominated in bitcoins.

\subsubsection{The \unit{\bsq} Coloured-Coin}\label{sec:bsq}

The Bisq \gls*{dao} operates by tracking the actions of a token or
coloured-coin, \unit{\bsq}, issued on the Bitcoin blockchain.
\unit{\bsq} tokens are simply transaction outputs that are coloured
and tracked by the Bisq software.  \unit{\bsq} transactions are
Bitcoin transactions, recognised as valid transactions by Bitcoin
nodes, and recognised as valid transactions by the Bisq software.
Participants of the Bisq \gls*{dao} must first hold some \unit{\bsq}
to make and vote upon proposals.  There is a two-sided market
for \unit{\bsq}.  On the supply side, \unit{\bsq} can be acquired in
several ways.  \unit{\bsq} was minted and distributed in a genesis
transaction on 15th April 2019.  Additionally, new \unit{\bsq} is
minted and distributed after each \gls*{dao} cycle to contributors
using the proposal and stake-based voting system.  \unit{\bsq} can
also be transacted between parties using transfer transactions.  On
the demand side, traders using Bisq can opt to pay trade fees at a
reduced rate by acquiring and burning \unit{\bsq}, thereby creating a
demand for \unit{\bsq} and rewarding contributors indirectly.  In this
way, \unit{\bsq} is used to financially reward contributors as well as
manage the operations of the Bisq \gls*{dao} itself.

Every action on the Bisq \gls*{dao}, such as a proposal or vote, takes
the form of a \unit{\bsq} transaction.  There are twelve transaction
types:

\begin{description}
  \item[Trade fee] transactions pay Bisq trade fees at a reduced
    rate using \unit{\bsq}.  The reduced rate incentivises users trading
    on Bisq to pay using \unit{\bsq} rather than bitcoin, thereby
    creating a demand for \unit{\bsq}\@.
  \item[Transfer] transactions transfer \unit{\bsq} between
    addresses in much the same way as Bitcoin transactions transfer
    bitcoin.
  \item[Compensation request] transactions request \unit{\bsq}
    compensation for contributions to the Bisq project.  Users supply
    non-coloured Bitcoin that will be coloured as \unit{\bsq} should the
    request be accepted by vote.  The details of the request are stored
    in an off-chain document such as a GitHub issue.  The transaction
    includes an \texttt{OP\_RETURN} containing the hash of the off-chain
    document.
  \item[Reimbursement request] transactions are functionally
    similar to compensation requests.  They reimburse users for
    out-of-pocket expenses relating to Bisq or compensate users for
    failed trades.
  \item[Proposal] transactions make proposals that are neither
    compensation nor reimbursement requests.  These include approving
    support agents and modifying trade fees.
  \item[Blind vote] transactions vote on open requests and
    proposals during the blind vote stage of a \gls*{dao} cycle.
  \item[Vote reveal] transactions publish unblinded votes during
    the vote reveal stage of a \gls*{dao} cycle.
  \item[Lockup] transactions lock \unit{\bsq} for a specified
    duration.  They are often used as a bond for a specified role in
    Bisq such as a trade mediator or arbitrator.
  \item[Unlock] transactions unlock previously locked
    \unit{\bsq}\@.
  \item[Asset listing fee] transactions list new tradable assets
    on Bisq, such as a new altcoin.  Assets are initially listed for a
    trial period.  If they do not reach a minimum trade volume, they are
    removed.
  \item[Proof of burn] transactions destroy \unit{\bsq}.  They
    do not have a specific use case but can be used as a form of
    reputation by proving that an individual burned \unit{\bsq}\@.
  \item[Genesis] the initial transaction
    that minted and distributed the initial quantity of \unit{\bsq}\@.
\end{description}

\begin{table}
  \centering
  \caption{The twelve valid \unit{\bsq} transaction types, their
    counts and whether or not they are
  self-transfers.}\label{tab:bsq-tx-types}
  \begin{tabular}{lrc}
    \toprule
    \multicolumn{1}{c}{} &
    \multicolumn{1}{c}{Count} &
    \multicolumn{1}{c}{Self-Transfer?}\\
    \midrule
    Trade Fee & \num{51785} & \cmark\\
    Transfer & \num{3489} & \xmark\\
    Compensation Request & \num{482} & \cmark\\
    Blind Vote & \num{368} & \cmark\\
    Vote Reveal & \num{365} & \cmark\\
    Proposal & \num{118} & \cmark\\
    Lockup & \num{48} & \cmark\\
    Proof of Burn & \num{37} & \cmark\\
    Reimbursement Request & \num{36} & \cmark\\
    Asset Listing Fee & \num{26} & \cmark\\
    Unlock & \num{17} & \cmark\\
    Genesis & \num{1} & \xmark\\
    \bottomrule
  \end{tabular}
\end{table}

\subsection{The Self-Transfer Issue}

Due to the Bisq \gls*{dao}'s reliance on the \unit{\bsq} token, a
significant amount of \gls*{dao} related activity is published to the
Bitcoin blockchain.  For example, if a participant makes a
compensation request, their Bisq node generates a \unit{\bsq}
transaction on the Bitcoin blockchain that is a compensation request
transaction.  Furthermore, by inspecting the specification of the
compensation request
transaction,\footnote{\url{https://docs.bisq.network/dao-technical-overview.html}}
we can see that the transaction inputs and the transaction outputs
always belong to the same participant.  In other words, the
participant is making a self-transfer with special data included in an
\texttt{OP\_RETURN} to signal to the Bisq \gls*{dao} that this is a
compensation request.  We can inspect the specifications of the
remaining transactions types, and identify the transactions that are
self-transfers.  Most transactions are
self-transfers.  In the list of twelve transaction types above, all
but the transfer transactions and the genesis transaction are
self-transfers.

This points to our Bisq \gls*{dao} address clustering heuristic: for
each self-transfer \unit{\bsq} transaction, the addresses referenced
by all of its transaction inputs and all of its transaction outputs
belong to the same participant; for each \unit{\bsq} transfer
transaction, the addresses referenced by all of its transaction inputs
and all but the first of its transaction outputs belong to the same
participant.  Only the address referenced by the first transaction
output in a \unit{\bsq} transfer transaction belongs to the recipient
rather than the sender.  The self-transfer issue allows the addresses
referenced at either side of these transactions to be clustered.  Only
the \unit{\bsq} genesis transaction and transfer transactions are not
necessarily self-transfers.  This information can be derived directly
from the specifications of the \unit{\bsq} transaction types.
\cref{app:dao-heuristic} provides pseudo-code for the heuristic and we
have released a C\#/.NET
implementation.\footnote{\url{https://github.com/Liam-Hickey-Ire/BisqDAOAnalysisSource}}

We have specified a heuristic by which the addresses associated with
\unit{\bsq} transactions can be clustered.  This is a heuristic
because it is possible for a participant to manually construct a
\unit{\bsq} transaction that violates these assumptions.  This could
cause our heuristic to generate misleading information.  It could
identify two or more addresses as belonging to a single entity, when
in fact they were controlled by separate entities.  Or, it could
identify two addresses as belonging to separate entities, when in fact
they were controlled by the same entity.  This is a perennial problem
with address clustering heuristics.  The assumptions underlying the
multi-input heuristic (see \cref{sec:related-work}) can be violated by
CoinJoin transactions.  This is the analogous situation for our
heuristic.  However, the ability to construct such transactions is not
supported by the Bisq software, that is, the only way to transfer
\unit{\bsq} is to create a \unit{\bsq} transfer transaction, and we
are not aware of any additional tools to support this.  We will
discuss the impact of false positives and false negatives in
\cref{sec:address-tagging} and \cref{sec:countermeasures}.

In this article we analyse all \num{56775} \unit{\bsq} transactions as
of Bitcoin block height \num{670026} after the completion of Bisq
\gls*{dao} Cycle \num{21} on 10th February
2021.\footnote{\url{https://github.com/Liam-Hickey-Ire/BisqDAOAnalysisSource/blob/master/Data/670026/dao-txes.csv}:
Each line contains the hash of a Bisq \gls*{dao} transaction.}
\cref{tab:bsq-tx-types} shows the distribution of the \unit{\bsq}
transaction types, excluding three irregular transactions.  We note
that \qty{91}{\percent} of the transactions burn \unit{\bsq} for trade
fees and \qty{94}{\percent} are self-transfers:  participants burn
\unit{\bsq} or signal an action to the Bisq \gls*{dao} (submitting
proposals, voting, or locking \unit{\bsq}), but the remaining
\unit{\bsq} and the underlying bitcoin are returned to the same
participant.

\section{Analysis and Results}\label{sec:analysis}

The transaction inputs and outputs of the \num{56775} \unit{\bsq}
transactions reference \num{163430} distinct addresses.\footnote{Our
  number differs from that shown on the \unit{\bsq} Block Explorer
  (\url{https://explorer.bisq.network}) since our number includes
addresses not carrying \unit{\bsq}-coloured bitcoins.} The address
clustering heuristic produces \num{1532} address
clusters.\footnote{\url{https://github.com/Liam-Hickey-Ire/BisqDAOAnalysisSource/blob/master/Data/670026/addr-clusters.csv}:
Each line contains the addresses in an address cluster.} That is, it
partitions the \num{163430} addresses into \num{1532} subsets such
that all addresses in the same subset are likely controlled by the
same participant.  The median number of addresses per address cluster
is $25$ ($\textrm{mean} = 107$, $\textrm{std.~dev.} = 274$).  This is
due to the Bisq software creating new addresses for each \unit{\bsq}
transaction.  Although this is good from a privacy perspective, it is
often negated by address clustering~\citep{harrigan-fretter-16}.

Generally, it is difficult to assess the validity of an address
clustering due to the unavailability of a ground
truth~\citep{nick-15}.  However, the Bisq \gls*{dao} offers the
following partial solution.  We assign a \emph{role} to each address
cluster:

\begin{enumerate}
  \item If an address cluster contains at least one address referenced
    by a transaction output of a \unit{\bsq} proposal transaction, we
    assign it the \emph{proposer} role.
  \item If an address cluster is not a proposer but it contains at least
    one address referenced by a transaction output of the \unit{\bsq}
    genesis transaction, we assign it the \emph{generator} role.
  \item If an address cluster is neither a proposer nor a generator, we
    assign it the \emph{user} role.
\end{enumerate}

Then, we can make use of address tagging to assess the validity of our
address clustering.

\subsection{Address Tagging}\label{sec:address-tagging}

There are \num{1244} users, \num{175} generators and \num{113}
proposers.  The roles are significant because we can assign
\emph{tags}, or links to pseudonyms and real-world identities, to all
of the proposers using data stored by the Bisq \gls*{dao} for the
\unit{\bsq} compensation, reimbursement and proposal transactions.
Furthermore, we can assign tags to many of the generators using GitHub
account usernames associated with transaction outputs of the
\unit{\bsq} genesis transaction.

Prior to the launch of the Bisq \gls*{dao} and the \unit{\bsq}
coloured-coin, the Bisq community performed the operations of the Bisq
\gls*{dao} and managed the issuance and circulation of prototypical
\unit{\bsq} coloured-coins manually and centrally.  During this
bootstrapping phase, the Bisq community tracked voting and staking
using a spreadsheet.\footnote{\url{https://long.af/kcaift}}
Additionally, contributors creating compensation requests at this time
stated the \unit{\bsq} address to which compensation should be
directed in the request's associated GitHub issue.  Using the
addresses found in both the spreadsheet and within the issues found on
GitHub, we created a pre-launch \unit{\bsq} tag database.

The Bisq \gls*{dao} was launched on the 15th April 2019.  \unit{\bsq}
holders were given the opportunity to specify the address they wished
to use in the \unit{\bsq} genesis transaction.  They could take one of
three actions:  retain their pre-launch address; publicly announce a
new address or change their address privately by notifying the
individual(s) who constructed the genesis transaction.  For each of
these cases, we can create a mapping from pre-launch addresses to
post-launch addresses, thus creating a post-launch tag database for
addresses referenced by the \unit{\bsq} genesis transaction.  Creating
a mapping for the first two cases is trivial as addresses are publicly
stated on
GitHub.\footnote{\url{https://github.com/bisq-network/compensation/issues/260}
and \url{https://github.com/bisq-network/compensation/issues/263}}
We were, however, also able to ascertain post-launch addresses for
those who chose to change their addresses privately.  We found that
the ordering of the transaction outputs of the \unit{\bsq} genesis
transaction matched the ordering of the entries in the spreadsheet.

Together, we can assign tags to \num{134} distinct address
clusters.\footnote{\url{https://github.com/Liam-Hickey-Ire/BisqDAOAnalysisSource/blob/master/Data/670026/cluster-tags.csv}:
Each line contains the tags assigned to an address cluster.} We stress
that assigning tags to individual addresses is trivial; the
information is publicly available and released by the proposers and
generators.  We, by contrast, are assigning tags to entire address clusters
generated using our Bisq \gls*{dao} address clustering heuristic
and all of their constituent activity, including trading, voting
and transfers.

Returning to the question of validity, we inspected the tags assigned
to each address cluster.  Out of the \num{134} tagged address
clusters, we identified nine with conflicting tags: nine address
clusters were assigned multiple tags that, ignoring obvious
capitalisation and spelling errors, were not the same.  This could be
an indication of false positives generated by our address clustering
heuristic.  However, on further inspection, we observe that at least
one case contains three different pseudonyms who submitted three
different \unit{\bsq} compensation proposal transactions for
overlapping translation contributions.  In many of the other cases we
observe real-world names combined with pseudonyms.  We do not believe
these are false positives but evidence of participants operating under
multiple aliases.  The privacy risk is stark.

Additionally, there are shared tags: several address clusters were
assigned tags that were identical to tags assigned to other address
clusters.  These are false negatives generated by our address
clustering heuristic.  They may be due to participants managing
multiple Bisq nodes with distinct \unit{\bsq} wallets or migrating
between \unit{\bsq} wallets using \unit{\bsq} transfer transactions.
We use the shared tags to reduce the number of address clusters to
\num{1504}\footnote{\url{https://github.com/Liam-Hickey-Ire/BisqDAOAnalysisSource/blob/master/Data/670026/addr-clusters-2.csv}:
Each line contains the addresses in an improved address cluster.} and
the number of tagged clusters to
\num{106}.\footnote{\url{https://github.com/Liam-Hickey-Ire/BisqDAOAnalysisSource/blob/master/Data/670026/cluster-tags-2.csv}:
Each line contains the tags assigned to an improved address cluster.}
In the context of address clustering, a false negative is less serious
than a false positive: assuming that two address clusters may be
controlled by two separate participants when in fact they are
controlled by one is a lack of information whereas assuming that one
address cluster is controlled by one participant when in fact it is
controlled by more than one is incorrect information.

\subsection{The Address Cluster Graph}

Once we have generated the address clusters, we can perform
higher-level analyses of activity within the Bisq \gls*{dao}\@.  We
can construct an \emph{address cluster graph} where each vertex
corresponds to an address cluster or Bisq \gls*{dao} participant and
each edge corresponds to a set of \unit{\bsq} transfer transactions
where the source and target vertices represent the sender and
recipient of the transactions, respectively.  \cref{fig:cluster-graph}
is a visualisation of the largest connected component of the address
cluster graph where the total value of the transactions associated
with each edge exceeds \qty{10000}{\bsq}\@.  This is an arbitrary
value chosen to produce a graph whose size is suitable for this
article; an interactive graph visualisation system is required to
navigate the entire graph.

The colour of each vertex represents the role of the corresponding
address cluster: red vertices are proposers; blue vertices are
generators and white vertices are users.  The size of each vertex is
proportional to the total amount of \unit{\bsq} sent to the addresses
in the corresponding address cluster.  We note that all of the red
vertices can be linked with pseudonyms, GitHub account names or
real-world names.

The address cluster graph represents a financial network where the
vertices represent Bisq \gls*{dao} participants, some of whom are
identifiable, and the edges represent financial relationships.
\cref{fig:cluster-graph} is a small subgraph of the address cluster
graph.  However, it points to the type of tools that can be built to
investigate Bisq activity.  For example, suppose a particular proposer
(red vertex) is the subject of an investigation but they have
carefully separated their real-world identity from their Bisq
activity.  They will appear in the address cluster graph without
immediately identifying information.  But it is a simple matter to
identify nearby vertices who have financially interacted with the
proposer in the past, and may possess identifying information about
the proposer.  Blockchain analysis service providers offer similar
functionality to identify Bitcoin users: they automatically point to
centralised services that may possess \emph{Personally Identifiable
Information} (PII) for users or their close
contacts~\citep{turner-et-al-20}.  This is a privacy risk since it
implies the applicability of a multitude of financial network analysis
techniques.

\begin{figure}
  \myincludegraphics[width=\linewidth]{graphml/address-clustering/bisq-cluster-graph}
  \caption{A graphical summary of the significant flows of \unit{\bsq}
    between address clusters (Bisq \gls*{dao} participants).  Vertex
    colour indicates role (red for proposers, blue for generators, and
    white for users) while vertex size indicates transaction volume;
  see the text for the full details.}
  \label{fig:cluster-graph}
\end{figure}

\subsection{The Two-Sided \unit{\bsq} Market}

All \unit{\bsq} originates with contributors to the Bisq project in
either the transaction outputs of the \unit{\bsq} genesis transaction
or the issuance transaction outputs of the accepted \unit{\bsq}
compensation and reimbursement request transactions.  Once minted,
\unit{\bsq} can be transferred between any number of participants
until it is eventually burnt, primarily by traders for trading fees.
We can use the address cluster graph to classify the \unit{\bsq}
transfer transactions based on the roles of the sender (the source
address cluster) and the recipient (the target address cluster).  The
breakdown for the \num{3489} \unit{\bsq} transfer transactions is
shown in \cref{tab:bsq-transfer-tx-types}.  For example, there are
\num{860} transfers from users to users, \num{20} transfers from users
to generators, and so on.  Although there are far fewer proposers
(\num{113}) and generators (\num{175}) than users (\num{1244}), the
proposers and generators are involved in \qty{75}{\percent} of all
\unit{\bsq} transfer transactions.

\begin{table}
  \centering
  \caption{We can classify the \num{3489} \unit{\bsq} transfer
    transactions based on the role of the sender and the receiver.
    The rows indicate the role of the sender and the columns indicate
  the role of the receiver.}\label{tab:bsq-transfer-tx-types}
  \begin{tabular}{lrrr}
    \toprule
    \multicolumn{1}{c}{} &
    \multicolumn{1}{c}{User} &
    \multicolumn{1}{c}{Generator} &
    \multicolumn{1}{c}{Proposer}\\
    \midrule
    User & \num{860} & \num{20} & \num{154}\\
    Generator & \num{126} & \num{15} & \num{35}\\
    Proposer & \num{1091} & \num{25} & \num{242}\\
    \bottomrule
  \end{tabular}
\end{table}

A similar situation presents itself in Bitcoin: large centralised
services such as exchanges, mining pools, gambling services and
darknet markets generate \textquote{super-clusters} in the address
clustering of the Bitcoin blockchain~\citep{harrigan-fretter-16}.
Even though they are few in number when compared with the total number
of Bitcoin users, they have high degree centrality in the
corresponding address cluster graph and are involved in a significant
number of Bitcoin transactions~\citep{lischke-fabian-16}.  Because of
this they are a focus of regulators and blockchain analysis service
providers.  Within Bisq, the proposers and generators could attract a
similar focus: they are involved in a significant number of
\unit{\bsq} transfer transactions, they play a central role in the
network and many of them are easily identifiable.

\subsection{The Dominant \unit{\bsq} Transactors}

At the time of our analysis, the Bisq \gls*{dao} had minted
\qty{5823351.09}{\bsq}, the participants had burnt
\qty{1323984.99}{\bsq}, primarily for trade fees, and
\qty{4499366.10}{\bsq} remained in circulation.  It is easy to
identify the address clusters that have transacted the most
\unit{\bsq}\@.  Out of the top ten \unit{\bsq} transactors, five can
be linked with GitHub account names and real-world names.  The
individuals are providing their names when submitting \unit{\bsq}
compensation and reimbursement proposal transactions.  Our address
clustering heuristic is linking this information with the entirety of
their Bisq \gls*{dao} activity including their transaction volume and
balances.

Decentralisation can result in governments placing increased
regulatory burdens on individuals.  For example, at the turn of the
century, \citet{newman-95} made the following observation in relation
to out-of-state sales taxes in the U.S.: \textquote{The irony of the
  movement toward local control and decentralising government is that
  the increased dependence on local taxes and revenue in an
  increasingly global retail market is pushing governments towards
  policies of more burdensome regulation on business and more
  intrusive government on the individual in order to collect those
  out-of-state sales taxes.  As local regions becoming increasingly
  artificial boundaries for government jurisdictions, even more
  jerry-rigged regulations are attempted to salvage regional financial
health.}

Decentralised exchanges, and those who participate in them, may suffer
a similar fate.  If the dominant \unit{\bsq} transactors can be
identified and linked with the entirety of their Bisq \gls*{dao}
activity including their transaction volume and balances, then they
may become a target for increased regulation.

\subsection{Impact Within the Bitcoin Blockchain}

We have assessed the Bisq \gls*{dao} and \unit{\bsq} token in
isolation.  However, all \unit{\bsq} transaction data is published to
the Bitcoin blockchain.  The set of \unit{\bsq} transactions is, by
definition, a subset of the set of Bitcoin transactions.  We can
assess the impact of the Bisq \gls*{dao} on address clusterings of the
entire Bitcoin blockchain.  We can also combine the Bisq Trade
Protocol address clustering heuristic
(\cref{sec:bisq-trade-protocol}), the Bisq \gls*{dao} address
clustering heuristic (\cref{sec:bisq-dao}), and the heuristics
referenced in \cref{sec:related-work}.  The address clusters generated
by our heuristic are equally valid when viewed through the lens of the
larger Bitcoin blockchain.  In fact, \unit{\bsq} addresses are simply
Bitcoin addresses with a leading \texttt{B} character.  By extension,
the observations stemming from the use of this heuristic are equally
applicable.

\section{Countermeasures}\label{sec:countermeasures}

There is no silver bullet to addressing the privacy leaks presented in
the previous sections.  Both the Bisq Trade Protocol heuristic
(\cref{sec:bisq-trade-protocol}) and the Bisq \gls*{dao} heuristic
(\cref{sec:bisq-dao}) extend the multi-input heuristic that is an
effective means of aggregating user activity on the Bitcoin
blockchain~\citep{harrigan-fretter-16}.  A coloured-coin system, such
as Bisq, that builds on top of Bitcoin will be subject to this type of
analysis.  Furthermore, the most obvious countermeasures involve
obfuscation through additional transactions, which incur transaction
fees.  This is undesirable since transaction fees on the Bitcoin
blockchain are expensive and expected to rise.

The Bisq Trade Protocol heuristic relies on the structure of the maker
fee, taker fee, deposit and payout transactions (see
\cref{fig:bisq-trade-protocol}).  We can follow the trade amount and
the security deposits through the four transactions.  The Bisq
software could obfuscate the flow of the security deposits by keeping
the security deposit belonging to the buyer of bitcoins separate from
the trade amount and randomising the order of the transaction outputs
containing the security deposits in the payout transaction.  However,
this would also necessitate careful coin-control by the buyer in
future transactions~\citep{ranvier-13,hearn-13}.

Several approaches can defeat the Bisq \gls*{dao}
heuristic.  The heuristic relies on \unit{\bsq} self-transfer
transactions.  The Bisq software could trigger false positives or
false negatives in this heuristic by introducing ambiguity into the
distinction between self-transfers and non-self-transfers.  Other than
the \unit{\bsq} genesis transaction, transfer transactions are the
only \unit{\bsq} transactions that are not entirely self-transfers.
As a result, transfer transactions have the effect of separating
clusters generated by our heuristic.  Disguising transfer transactions
so that they cannot be distinguished from self-transfer transactions
would trigger false positives in the heuristic, invalidating generated
clusters.  For example, a participant could create a \unit{\bsq} trade
fee transaction to transfer \unit{\bsq} where the \textquote{change}
was directed to the recipient and a small amount of \unit{\bsq} was
burnt to satisfy the requirement of a \unit{\bsq} trade fee
transaction.  While this solution defeats the heuristic as it stands,
there are other ways in which \unit{\bsq} transaction types can be
deduced.  Every trade fee transaction can be linked to the four
transactions (maker fee, taker fee, deposit and payout) of a Bisq
trade.  Consequently, any trade fee transaction that is not linked to
a Bisq trade could be identified as a disguised transfer transaction
and treated as such.

Additionally, transfer transactions can be used to trigger false
negatives in our heuristic, thereby diminishing the heuristic's
effectiveness.  Triggering a false negative requires the use of
\textquote{dummy} transfer transactions after each self-transfer
transaction.  This transfer transaction sends \unit{\bsq} from the
change address used in the last self-transfer to a new address owned
by the same user.  This gives the appearance of \unit{\bsq} being sent
between parties, thus reducing the size of the address clusters
generated by our heuristic.  While dummy transfer transactions reduce
the effectiveness of the heuristic, they also create transactions that
are not otherwise needed, increasing the cost for users.  Of course,
functionality to create dummy transactions and a best-practices guide
could be included in the Bisq software and documentation and only used
to improve privacy as required.

The Bisq Trade Protocol heuristic and the Bisq \gls*{dao} heuristic
rely on the ordering of transaction inputs and transaction outputs of
particular transactions created by the Bisq software, see
\cref{app:trade-heuristic,app:dao-heuristic}.  In the case of the Bisq
Trade Protocol, the orderings of the transaction inputs and
transaction outputs of the four transactions (maker, taker, deposit
and payout) could be randomised if both peers agreed on the orderings
as part of each Bisq trade contract.  However, the orderings could
still be deduced by following the trade amount from the seller to the
buyer.  Similarly, the Bisq \gls*{dao} could randomise the ordering of
the transaction outputs of a \unit{\bsq} transfer transaction that
carry \unit{\bsq}.  The first transaction output carrying \unit{\bsq}
is for the recipient of the transfer.  The second transaction output
carrying \unit{\bsq}, if it exists, is change for the sender.
Even if their ordering were randomised on a per-transfer
basis, we could still use a change heuristic to tell them apart, see,
for example,
\citet{meiklejohn-et-al-16}.

\citet{haveno-xx} is a fork of Bisq based on Monero~\citep{serhack-18}
instead of Bitcoin with improved privacy.  It uses a one-time ring
signature to weaken traceability: when signing a transaction, the
signer can search for other entries in the Unspent Transaction Output
Set (\glstext*{utxo}) and create a one-time ring signature that could have been
created by the owners of any of those transaction outputs.  This
renders the Bisq Trade Protocol heuristic ineffective.  Haveno has
also removed the Bisq \gls*{dao} completely.  Traders must pay for
trade fees using Monero's native currency (\unit{\xmr}).  However, the
Bisq \gls*{dao} and the \unit{\bsq} token transfer value from the
traders using Bisq to the contributors maintaining it,
without requiring a centralised treasury.  It is unclear how the
Haveno treasury will be managed and whether it will encourage community
participation.

\section{Conclusions}\label{sec:conclusions}

We demonstrated the privacy cost of participating in Bisq trades and
the Bisq \gls*{dao}\@.  First, we showed that Bisq trades on the
Bitcoin blockchain can be easily identified.  The trade amount in
bitcoins and the security deposits can be linked with their respective
owners before and after a trade.  This points to a Bisq Trade Protocol
address clustering heuristic.  Second, participants of the Bisq
\gls*{dao} disclose significant information about themselves,
especially when submitting \unit{\bsq} compensation and reimbursement
proposal transactions.  Even though Bisq generates new addresses for
every \unit{\bsq} transaction, \qty{94}{\percent} of these
transactions are self-transfers, that is, all of the transaction inputs
and outputs belong to the same participant.  This points to a Bisq
\gls*{dao} address clustering heuristic.  We implemented this
heuristic and applied it to all \unit{\bsq} transactions to date.  The
heuristic proves effective in aggregating all activity performed by
each participant such as trades, votes and proposals.  We can attach
pseudonyms, GitHub account names and real-world names to many of the
central participants.  This has important implications for user
privacy.  Although not examined in this article, it has further
implications for the Bisq \gls*{dao} voting system and address
clustering in the broader Bitcoin blockchain.

The Bisq Trade Protocol and the Bisq \gls*{dao} are innovative
approaches to peer-to-peer trading of bitcoins for other currencies
and to decentralising the governance and finance functions of a
decentralised exchange.  However, when viewed through the prism of
blockchain analysis and address clustering, they appear vulnerable.
Traders and participants of the Bisq \gls*{dao} will expect certain
limits on what is known about them and on what others can find out.
Blockchain analysis could unsettle this expectation and have a
\textquote{chilling effect} on adoption.

\section*{Acknowledgements}

The authors are grateful to the members of the Bisq community and the
anonymous reviewers that provided feedback on earlier versions of this
article.

\bibliographystyle{my-abbrvnat}
\bibliography{bisq}

\section*{Appendices}

\appendix

\section{The Bisq Trade Protocol Address Clustering
Heuristic}\label{app:trade-heuristic}

We identify all Bisq Trade Protocol transactions on the Bitcoin
blockchain using their distinctive structure (see
\cref{sec:bisq-trade-protocol} and \cref{fig:bisq-trade-protocol}).
Once we have identified the trades, we iterate through each one and
cluster the associated addresses.  \textsc{ClusterBisqTrade} clusters
the addresses associated with a single Bisq trade (see
\cref{alg:trade-heuristic}).  The inputs are the maker fee transaction
($M$), the taker fee transaction ($T$) and the payout transaction
($P$).  $P$ always has exactly two transaction outputs:  the first
contains the security deposit for the seller; the second contains the
security deposit and the trade amount for the buyer.  The relevant
addresses are the addresses belonging to the transaction inputs of $M$
(Line~\ref{lin:maker-fee-inputs}), the addresses belonging to the
transaction outputs of $M$ (Line~\ref{lin:maker-fee-outputs}), the
addresses belonging to the transaction inputs of $T$
(Line~\ref{lin:taker-fee-inputs}), the addresses belonging to the
transaction outputs of $T$ (Line~\ref{lin:taker-fee-outputs}), and the
addresses belonging to the transaction outputs of $P$
(Line~\ref{lin:payout-outputs}).  The output is a partition of the
addresses into subsets such that all addresses in the same subset are
controlled by the same entity.

\textsc{IsSeller?} distinguishes between two cases: either $M$ is the
seller or $T$ is the seller (Line~\ref{lin:two-cases}).  $M$ and $T$
always have at least two transaction outputs where the second
transaction output contains either a security deposit or a security
deposit combined with the trade amount.  In addition, the second
transaction output of the seller is spent by the second transaction
input of $D$.  The ordering of the transaction inputs and transaction
outputs can be deduced from the Bisq source
code.\footnote{\url{https://github.com/bisq-network/bisq/blob/7233979d94abde020eadaab7dae33b0efb0e2e7e/core/src/main/java/bisq/core/btc/wallet/TradeWalletService.java}:
Line~304} Therefore, by inspecting $M$ and $D$, we can deduce if the
maker is the seller or the taker is the seller.  If the maker is the
seller then we cluster the addresses belonging to the transaction
inputs and outputs of $M$ with the address belonging to the first
transaction output of $P$, and we cluster the addresses belonging to
the transaction inputs and outputs of $T$ with the address belonging
to the second transaction output of $P$ (Line~\ref{lin:case-one}).
Alternatively, if the taker is the seller we cluster the addresses
belonging to the transaction inputs and outputs of $M$ with the
address belonging to the second transaction output of $P$, and we
cluster the addresses belonging to the transaction inputs and outputs
of $T$ with the address belonging to the first transaction output of
$P$ (Line~\ref{lin:case-two}).  In both cases we are simply following
the flow of bitcoins from a seller to a buyer.

We note that the addresses belonging to the first transaction outputs
of both $M$ and $T$, and the addresses belonging to the transaction
input of $P$, are not clustered.  The input of $P$ is only used by the
Bisq Trade Protocol itself and is never reused, while the first output
of both $M$ and $T$ contain Bisq trade fees.

\begin{algorithm}[H]
  \caption{The Bisq Trade Protocol Address Clustering Heuristic}
  \label{alg:trade-heuristic}
  \begin{algorithmic}[1]
    \Function {ClusterBisqTrade}{$M$, $T$, $D$, $P$}
    \State $\{a_0, a_1, \ldots, a_p\} \gets
    \Call{TxInAddrs}{M}$\label{lin:maker-fee-inputs}
    \State $\{b_0, b_1, b_2\} \gets
    \Call{TxOutAddrs}{M}$\label{lin:maker-fee-outputs}
    \State $\{c_0, c_1, \ldots, c_q\} \gets
    \Call{TxInAddrs}{T}$\label{lin:taker-fee-inputs}
    \State $\{d_0, d_1, d_2\} \gets
    \Call{TxOutAddrs}{T}$\label{lin:taker-fee-outputs}
    \State $\{e_0, e_1\} \gets \Call{TxOutAddrs}{P}$\label{lin:payout-outputs}
    \If{$\Call{IsSeller?}{M, D}$}\label{lin:two-cases}
    \State \textbf{return} $\{\{a_0, a_1, \ldots, a_p, b_1, b_2,
    e_0\}, \{c_0, c_1, \ldots, c_q, d_1, d_2, e_1\}\}$\label{lin:case-one}
    \Else
    \State \textbf{return} $\{\{a_0, a_1, \ldots, a_p, b_1, b_2,
    e_1\}, \{c_0, c_1, \ldots, c_q ,d_1, d_2, e_0\}\}$\label{lin:case-two}
    \EndIf
    \EndFunction
  \end{algorithmic}
\end{algorithm}

\section{The Bisq \glstext*{dao} Address Clustering
Heuristic}\label{app:dao-heuristic}

We identify and categorise all \unit{\bsq} transactions on the Bitcoin
blockchain (see \cref{sec:bisq-dao} and \cref{tab:bsq-tx-types}).
Once we have identified the transactions, we iterate through each one
and cluster the associated addresses.
\textsc{ClusterBisqDAOTransaction} clusters the addresses associated
with a single Bisq \gls*{dao}, or \unit{\bsq} transaction (see
\cref{alg:dao-heuristic}).  The only input is a \unit{\bsq}
transaction ($T$).  The relevant addresses are the addresses belonging
to the transaction inputs of $T$ (Line~\ref{lin:tx-inputs}) and the
addresses belonging to the transaction outputs of $T$
(Line~\ref{lin:tx-outputs}).  The output is a partition of the
addresses into subsets such that all addresses in the same subset are
controlled by the same entity.

\textsc{IsGenesisTx?} and \textsc{IsTransferTx?} distinguish between
three cases: either $T$ is the \unit{\bsq} genesis transaction, $T$ is
a \unit{\bsq} transfer transaction, or $T$ is some other type of
\unit{\bsq} transaction (see \cref{tab:bsq-tx-types}).  In the first
case, we cluster the addresses belonging to the transaction inputs of
$T$ (Line~\ref{lin:dao-case-one}) only.  Those addresses are
controlled by the creator of the \unit{\bsq} genesis transaction.
This information can be derived directly from the specification of the
genesis transaction.  In the second case, we cluster the addresses
belonging to the transaction inputs of $T$ and the addresses belonging
to all but the first transaction output of $T$.  Those addresses are
controlled by the creator of the \unit{\bsq} transfer transaction.
This information can also be derived directly from the specification
of a transfer transaction.  The address belonging to the first
transaction output of $T$ is the address of the recipient.  The
ordering of the transaction outputs can be deduced from the Bisq
source
code.\footnote{\url{https://github.com/bisq-network/bisq/blob/7233979d94abde020eadaab7dae33b0efb0e2e7e/core/src/main/java/bisq/core/btc/wallet/BsqWalletService.java}:
Line~565} In the third case, we cluster the addresses belonging to the
transaction inputs of $T$ and the addresses belonging to the
transaction outputs of $T$.  Those addresses are controlled by the
creator of the \unit{\bsq} transaction.  Again, information can be
derived directly from the specification of the remaining transaction
types.

\begin{algorithm}[H]
  \caption{The Bisq \gls*{dao} Address Clustering Heuristic}
  \label{alg:dao-heuristic}
  \begin{algorithmic}[1]
    \Function {ClusterBisqDAOTransaction}{$T$}
    \State $\{a_0, a_1, \ldots, a_p\} \gets
    \Call{TxInAddrs}{T}$\label{lin:tx-inputs}
    \State $\{b_0, b_1, \ldots, b_q\} \gets
    \Call{TxOutAddrs}{T}$\label{lin:tx-outputs}
    \If{$\Call{IsGenesisTx?}{T}$}
    \State \textbf{return} $\{\{a_0, a_1, \ldots, a_p\}, \{b_0\},
    \{b_1\}, \ldots, \{b_q\}\}$\label{lin:dao-case-one}
    \ElsIf{$\Call{IsTransferTx?}{T}$}
    \State \textbf{return} $\{\{a_0, a_1, \ldots, a_p, b_1, \ldots,
    b_q\}, \{b_0\}\}$\label{lin:dao-case-two}
    \Else
    \State \textbf{return} $\{\{a_0, a_1, \ldots, a_p, b_0, b_1,
    \ldots, b_q\}\}$\label{lin:dao-case-three}
    \EndIf
    \EndFunction
  \end{algorithmic}
\end{algorithm}

\end{document}